\documentclass[aps,twocolumn,floatfix]{revtex4}
\usepackage{graphicx}
\usepackage{amsmath}
\usepackage{color}
\newcommand{\be}{\begin{equation}}
\newcommand{\ee}{\end{equation}}
\newcommand{\bea}{\begin{eqnarray}}
\newcommand{\eea}{\end{eqnarray}}
\newcommand{\bs}{\begin{split}}
\newcommand{\bse}{\begin{subequations}}
\newcommand{\ese}{\end{subequations}}
\newcommand{\ecp}{${\rm EuCo_2P_2}$}

\begin{document}
 
\title{Magnetic structure and magnetization of $z$-axis helical Heisenberg antiferromagnets with XY~anisotropy in high  magnetic fields transverse to the helix axis at zero~temperature}
\author {David C.\ Johnston} 
\affiliation {Ames Laboratory and Department of Physics and Astronomy, Iowa State University, Ames, Iowa 50011}

\date{\today}

\begin{abstract}

A helix has a wavevector along the $z$ axis with the magnetic moments ferromagnetically-aligned within $xy$ planes with a turn angle $kd$ between the moments in adjacent planes in transverse field \mbox{${\bf H} = H_x\hat{\bf i} = 0$}.  The magnetic structure and $x$-axis average magnetization per spin of this system in a classical XY anisotropy field $H_{\rm A}$ is studied versus $kd$, $H_{\rm A}$, and large~$H_x$ at zero temperature.  For values of $H_{\rm A}$ below a \mbox{$kd$-dependent} maximum value, the $xy$~helix phase transitions with increasing $H_x$ into a spin-flop (SF) phase where the ordered moments have $x$, $y$, and $z$ components.  The moments in the SF phase are taken to be distributed on either one or two $xyz$~spherical ellipses.  The minor axes of the ellipses are oriented along the $z$~axis and the major axes along the $y$~axis where the ellipses are flattened along the $z$ axis due to the presence of the XY anisotropy.  From energy minimization of the SF spherical ellipse parameters for given values of $kd$, $H_{\rm A}$ and $H_x$, four $kd$-dependent SF phases are found: either one or two $xyz$ spherical ellipses and either one or two $xy$~fans, in addition to the $xy$~helix phase and the paramagnetic (PM) phase with all moments aligned along {\bf H}.  The PM phase occurs via second-order transitions from the $xy$~fan and SF phases with increasing $H_x$.  Phase diagrams in the $H_x$-$H_{\rm A}$ plane are constructed by energy minimization with respect to the SF phases, the $xy$~helix phase, and the $xy$~fan phase for four $kd$ values. One of these four phase diagrams is compared with the magnetic properties found experimentally for the model helical Heisenberg antiferromagnet \ecp\ and semiquantitative agreement is found.

\end{abstract}
\maketitle

\section{Introduction}

\begin{figure}[t]
\includegraphics [width=1.75in]{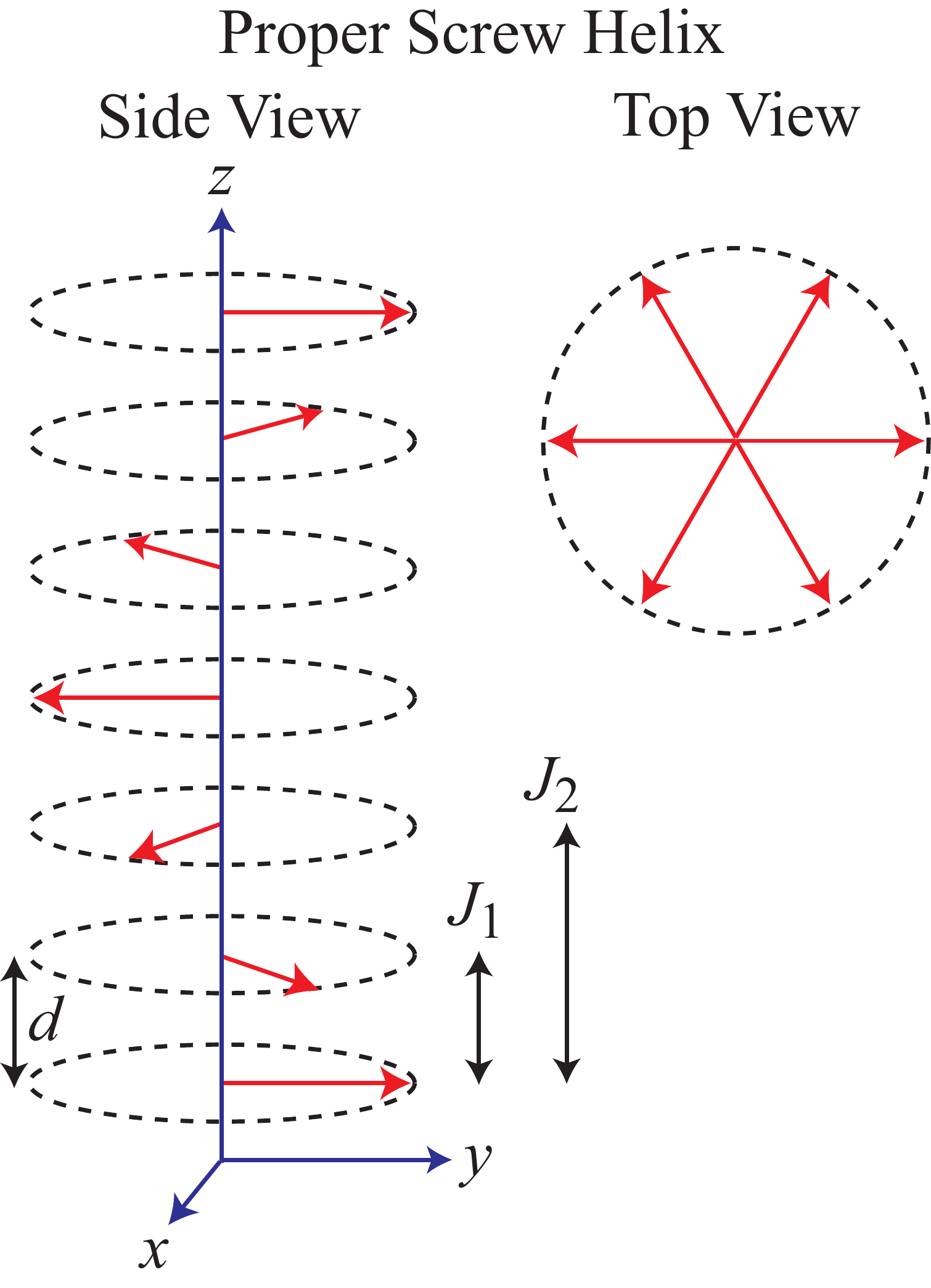}
\caption {Generic helical AFM structure \cite{Johnston2012}.  Each arrow represents a layer of moments perpendicular to the $z$~axis that are ferromagnetically aligned within the $xy$ plane and with interlayer separation $d$.  The wave vector {\bf k} of the helix is directed along the $z$~axis.  The magnetic moment turn angle between adjacent magnetic layers is $kd$.  The nearest-layer and next-nearest-layer exchange interactions $J_{1}$ and $J_{2}$, respectively, within the $J_0$-$J_{1}$-$J_{2}$ Heisenberg MFT model are indicated.  The top view is a hodograph of the magnetic moments.}
\label{Fig:J0_Jz1_Jz2_model_helix}
\end{figure}

A reformulation of the Weiss molecular field theory for Heisenberg magnets containing identical crystallographically-equivalent spins was developed recently, termed the unified molecular field theory (MFT), which treats collinear and noncollinear antiferromagnets on the same footing~\cite{Johnston2012, Johnston2015, Johnston2015b}.  The influences of magnetic-dipole and single-ion anisotropies  and classical anisotropy fields on the magnetic properties of such Heisenberg antiferromagnets were also studied within unified MFT~\cite{Johnston2016, Johnston2017, Johnston2017b}.

\begin{figure}
\includegraphics [width=2in]{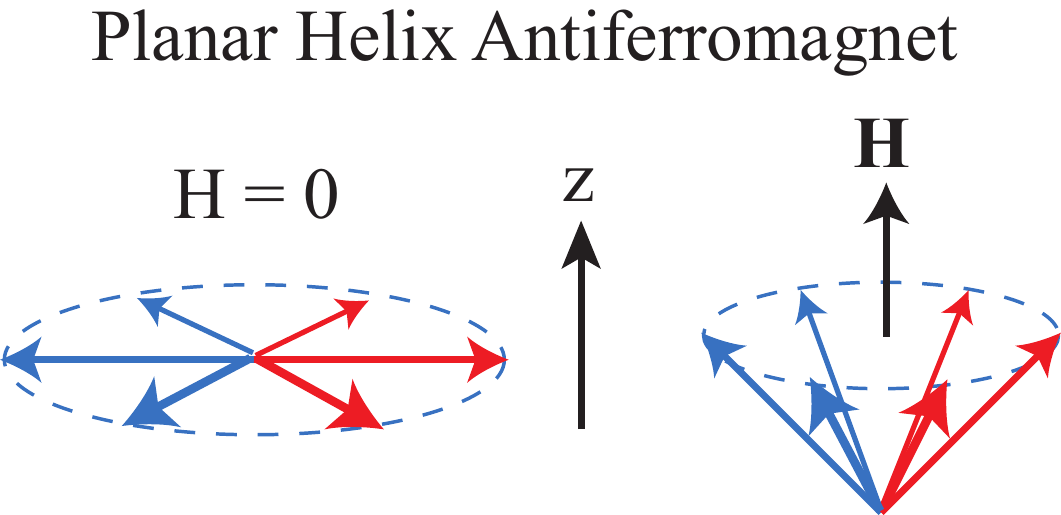}
\caption {Hodograph of the magnetic moments in a planar helical structure in the $xy$~plane with applied field (left figure)~$H=0$, and~(right figure) with a large ${\bf H}$ applied along the helix wave-vector $z$~axis \cite{Johnston2015}.}
\label{Fig:High_Perp_Field_Struct}
\end{figure}

Previously, the magnetic structure and magnetization of a planar helical antiferromagnet in a high applied magnetic fields~{\bf H} perpendicular to the helix wave vector axis ($z$~axis) at temperature $T=0$ was calculated where the ordered magnetic moments were restricted to lie in the $xy$~plane \cite{Nagamiya1962, Johnston2017c}.  This is the plane in which the ordered moments reside in zero field as shown in Fig.~\ref{Fig:J0_Jz1_Jz2_model_helix}.  This situation corresponds to infinite XY planar anisotropy.  Continuous crossover, second-order, and first-order transitions were found between the planar helix and planar fan phases with increasing $H$~\cite{Nagamiya1962, Johnston2017c}, the nature of which depends on the helix wave vector~$k$.  The influence of a high $z$-axis field on the magnetic moment vectors for the helix phase is shown in Fig.~\ref{Fig:High_Perp_Field_Struct}.  The magnetization versus field for this case was calculated in Ref.~\cite{Johnston2015}.  In Ref.~\cite{Johnston2017c}, the experimental high $xy$-plane field data at low temperatures~$T$ for a single crystal of the helical antiferromagnet \ecp~\cite{Sangeetha2016} containing Eu$^{+2}$ spins $S=7/2$ were fitted rather well by the theory for $k = 6\pi/7$, close to the value from neutron-diffraction measurements~\cite{Reehuis92}.  However, the presence of a field-induced out-of-plane component of the magnetic moments was not ruled out.

The $T=0$ calculations were extended to the case of finite XY anisotropy for fields applied perpendicular to the helix axis, where phase transitions between the helix, a three-dimensional spherical ellipse spin-flop (SF), $xy$ fan and the paramagnetic (PM) phases were found for small turn angles~$kd$ \cite{Nagamiya1962}.  Here we extend these $T=0$ calculations to arbitrary $kd$ values for finite classical XY anisotropy using our formulation of the classical XY anisotropy field ${\bf H}_{\rm A}$ within unified molecular field theory~\cite{Johnston2017b}.  We assume that for {\bf H} aligned along the $x$~axis, transverse to the helix $z$~axis, the moments can exhibit a transition to one of two types of three-dimensional SF spherical-ellipse phases with increasing $H_x$ with the $x$~axis intersecting the center of each spherical ellipse.  One type arises for either ferromagnetic (FM, $J_1<0$) or antiferromagnetic (AFM, $J_1>0$) nearest-layer interactions $J_1$ in Fig.~\ref{Fig:J0_Jz1_Jz2_model_helix} and the second type sometimes occurs for AFM $J_1$ at low $H_x$ and small $H_{\rm A1}$.  All helices have AFM $J_2>0$~\cite{Johnston2015}. The spherical-ellipse nature of the magnetic structures in the SF phase arises from the XY anisotropy and the fixed magnitude of the moments at $T=0$.

The average energy per spin of a helical spin system with the moments aligned in the $xy$~plane versus $H_x$ in the case of infinite XY anisotropy field $H_{\rm A}$ was calculated for $T=0$ in Ref.~\cite{Johnston2017c}.  Here we calculate the average energy per spin at finite $H_{\rm A}$, minimized at fixed~$kd$ and $H_{\rm A}$ with respect to the spherical ellipse parameters for the two types of spherical-ellipse SF phases and compare its energy at each field with that of the planar $xy$ helix/$xy$~fan phases at the same $H_x$ to determine the stable phase.  The PM phase arises naturally from the SF $\to xy$~fan~$\to$~PM and  $xy$~helix $\to$ $xy$~fan~$\to$~PM phase progression with increasing $H_x$.  This allows the magnetic phase diagram in the $H_x$--$H_{\rm A}$ plane at $T=0$ to be constructed, which we carry out for four values of the turn angle $kd$.  As part of these calculations, we obtain and present the \mbox{$x$-axis} average magnetic moment per spin $\mu_{x\,{\rm ave}}$ versus $H_x$ and $H_{\rm A}$ for the same four values of $kd$ which also reveal the phase transitions as well as their first- or second-order nature.

The unified MFT used in the present work is described in Sec.~\ref{Sec:Theory}, where the general aspects of the theory are reviewed in Sec.~\ref{Sec:GenThy} and the application of those to the one-dimensional $J_0$-$J_1$-$J_2$ model (see Fig.~\ref{Fig:J0_Jz1_Jz2_model_helix}) is given in Sec.~\ref{Sec:J0J1J2}.  The results for the SF phase are presented in Sec.~\ref{Sec:SFPhase}.  From minimization of the energy with respect to the SF, $xy$~helix, and $xy$~fan phases for four values of $kd$, the four resulting $T=0$ phase diagrams in the $H_x$-$H_{\rm A}$ plane are presented in Sec.~\ref{Sec:Phase Diagrams}, where our previous calculations for the energies of the $xy$~helix and $xy$~fan phases in Ref.~\cite{Johnston2017c} are utilized.  The methods needed to interface our theoretical $T=0$ phase diagrams with experimental low-$T$ magnetization versus field isotherms and magnetic susceptibility measurements versus~$T$ for helical Heisenberg antiferromagnets are presented in Secs.~\ref{Sec:hA1**hx**Convert} and~\ref{Sec:hA1etcExtraction}.  A comparison of the phase diagram for $kd=3\pi/4$~rad with the properties obtained from magnetic data for  \ecp\ with $kd = 0.85\pi$~rad~\cite{Sangeetha2016} is given in Sec.~\ref{Sec:EuCo2P2}, and semiquantitative agreement is found.  The results of the paper are summarized and discussed in Sec.~\ref{Sec:Summary}.

\section{\label{Sec:Theory} Theory}

\subsection{\label{Sec:GenThy} General theory}

All spins are assumed to be identical and crystallographically equivalent which means that they each have the same magnetic environment.  The magnetic moment $\vec{\mu}_n$ of spin~$n$ is
\bse
\label{Eqs:muvec}
\be
\vec{\mu}_n = -g\mu_{\rm B}{\bf S}_n 
\label{Eq:vecmun}
\ee
where the negative sign arises from the negative charge on an electron, $g$ is the spectroscopic splitting factor of each moment, $\mu_{\rm B}$ is the Bohr magneton, and ${\bf S}_n$ is the spin angular momentum of $\vec{\mu}_n$ in units of $\hbar$ which is Planck's constant~$h$  divided by~$2\pi$.  One can also write
\be
\vec{\mu}_n = \mu\,{\hat{\mu}}_n,
\label{Eq:mun}
\ee
where $\mu = |\vec{\mu}|$. At $T=0$ as considered in this paper, $\mu$ is the saturation moment given from Eq.~(\ref{Eq:vecmun}) as
\be
\mu = g\mu_{\rm B}S.
\label{Eq:muDef}
\ee
In Cartesian coordinates, the unit vector $\hat{\mu}_n$ in the direction of $\vec{\mu}_n$ is written as
\be
\hat{\mu}_n = \bar{\mu}_{nx}\,\hat{\bf i} + \bar{\mu}_{ny}\,\hat{\bf j} + \bar{\mu}_{nz}\,\hat{\bf k},
\label{Eq:hatmun}
\ee
where the Cartesian unit vectors pointing towards the positive $x,\ y$, and $z$~directions are $\hat{\bf i},\ \hat{\bf j}$ and~$\hat{\bf k}$, respectively, and 
\be
\bar{\mu}_{nx,ny,nz} \equiv \frac{\mu_{nx,ny,nz}}{\mu}.
\ee
Therefore
\be
\hat{\mu}_n \cdot \hat{\mu}_n = 1 = \bar{\mu}_{nx}^2 + \bar{\mu}_{ny}^2 + \bar{\mu}_{nz}^2.
\label{Eq:munDotmun}
\ee
\ese

The energy per spin $E_n$ of a representative spin~${\bf S}_n$ interacting with its neighbors ${\bf S}_{n^\prime}$ and with the classical anisotropy field ${\bf H}_{{\rm A}n}$ and applied magnetic field {\bf H} is
\be
E_n = E_{{\rm exch}n} + E_{{\rm A}n} + E_{Hn}.
\label{Eq:Enpieces}
\ee
The Heisenberg exchange energy per spin $E_{{\rm exch}n}$ is~\cite{Johnston2015}
\be
E_{{\rm exch}n} = \frac{1}{2}{\bf S}_n\cdot \sum_{n^\prime}J_{nn^\prime}{\bf S}_{n^\prime},
\label{Eq:En1}
\ee
where the prefactor of 1/2 is due to the fact that the exchange energy from interaction between a pair of spins is equally shared between the members of the pair, and $J_{nn^\prime}$ is the Heisenberg exchange interaction between spins ${\bf S}_n$ and ${\bf S}_{n^\prime}$.  Writing the classical expression
\be
{\bf S}_n\cdot {\bf S}_{n^\prime} = S^2\cos\alpha_{nn^\prime},
\ee
where $\alpha_{nn^\prime}$ is the angle between $\vec{\mu}_n$ and $\vec{\mu}_{n^\prime}$, Eq.~(\ref{Eq:En1}) becomes
\be
E_{{\rm exch}n} = \frac{S^2}{2}\sum_{n^\prime}J_{nn^\prime}\cos\alpha_{nn^\prime}.
\label{Eq:En9}
\ee
In terms of the magnetic moments, this can be written
\be
E_{{\rm exch}n} = \frac{S^2}{2}\sum_{n^\prime}J_{nn^\prime}\hat{\mu}_n\cdot \hat{\mu}_{n^\prime}.
\label{Eq:En19}
\ee

The anisotropy energy $E_{{\rm A}n}$ is assumed to arise from a classical anisotropy field ${\bf H}_{{\rm A}n}$ originating fundamentally from two-spin interactions (i.e., not from single-ion anisotropy) that is given by~\cite{Johnston2017b}
\be
E_{{\rm A}n} = -\frac{1}{2}\vec{\mu}_n\cdot {\bf H}_{{\rm A}n} = -\frac{\mu}{2}\hat{\mu}_n \cdot {\bf H}_{{\rm A}n},
\label{Eq:EA1n}
\ee
where the prefactor of 1/2 arises for the same reason as in Eq.~(\ref{Eq:En1}). The ${\bf H}_{{\rm A}n}$ seen by $\vec{\mu}_n$ is proportional to the projection of $\hat{\mu}_n$ onto the $xy$~plane according to~\cite{Johnston2017b}
\be
{\bf H}_{{\rm A}n} = \frac{3H_{\rm A1}}{S+1} \big(\bar{\mu}_{nx}\,\hat{\bf i} + \bar{\mu}_{ny}\,\hat{\bf j}\big),
\label{Eq:HAn}
\ee
where $H_{\rm A1}$ is the so-called fundamental anisotropy field.  Inserting Eqs.~(\ref{Eq:hatmun}) and~(\ref{Eq:HAn}) into~(\ref{Eq:EA1n}) and using Eq.~(\ref{Eq:muDef}) gives
\bea
E_{{\rm A}n} &=&  -\frac{3S}{2(S+1)} g\mu_{\rm B}H_{\rm A1}(\bar{\mu}_{nx}^2 + \bar{\mu}_{ny}^2)\nonumber\\
&=&-\frac{3S}{2(S+1)} g\mu_{\rm B}H_{\rm A1}(1 - \bar{\mu}_{nz}^2),\label{Eq:EA1n2}
\eea
where the second equality was obtained using Eq.~(\ref{Eq:munDotmun}).

The Zeeman energy $E_{Hn}$ of $\vec{\mu}_n$ in the applied magnetic field {\bf H} is
\be
E_{Hn} = -\vec{\mu}_n\cdot {\bf H}  = -\mu\bar{\mu}_{nx}H_x = -g\mu_{\rm B}S\bar{\mu}_{nx}H_x,
\label{Eq:EHn}
\ee
where Eqs.~(\ref{Eq:muDef}) and~(\ref{Eq:hatmun}) were used and {\bf H} is assumed to be applied in the $\hat{\bf i}$ direction, transverse to the helix $z$~axis, i.e.,
\be
{\bf H} = H_x \hat{\bf i}.
\label{Eq:HvecDef}
\ee

Inserting Eqs.~(\ref{Eq:En19}), (\ref{Eq:EA1n2}), and~(\ref{Eq:EHn}) into~(\ref{Eq:Enpieces}) gives the energy per spin as
\bea
E_n &=&  \frac{S^2}{2}\sum_{n^\prime}J_{nn^\prime}\hat{\mu}_n\cdot \hat{\mu}_{n^\prime}\nonumber\\ 
&& -\ \frac{3S}{2(S+1)} g\mu_{\rm B}H_{\rm A1}(1 - \bar{\mu}_{nz}^2) \label{Eq:EnTotal}\\
&& -\ \bar{\mu}_{nx} Sg\mu_{\rm B}H_x. \nonumber
\eea

\subsection{\label{Sec:J0J1J2} $\bf J_0$-$\bf J_1$-$\bf J_2$ one-dimensional MFT model for the exchange energy of helical antiferromagnets}

The $J_0$-$J_1$-$J_2$ unified MFT model for the Heisenberg exchange interactions~\cite{Johnston2012, Johnston2015} is utilized to treat helical structures such as illustrated in Fig.~\ref{Fig:J0_Jz1_Jz2_model_helix}, where $J_0$ is the sum of all Heisenberg exchange interactions between a representative spin~${\bf S}_n$ in a FM-aligned layer with all other spins in the same layer, $J_1$ is the sum of the interactions of that spin with all spins in a nearest-neighbor layer, and $J_2$ is the sum of the interactions of that spin with all spins in a next-nearest-neighbor layer, as shown in Fig.~\ref{Fig:J0_Jz1_Jz2_model_helix}.  Within this MFT model, the exchange energy of a representative spin ${\bf S}_n$ with magnitude~$S$ interacting with its neighbors is given by Eq.~(\ref{Eq:En9}) for $H_x = 0$ and with spins confined to the $xy$ plane as
\be
E_{{\rm exch}n} = \frac{S^2}{2}\left[J_0 + 2J_1\cos(kd) + 2 J_2 \cos(2kd)\right],
\label{Eq:EnJ0J1J2}
\ee
where $J_{nn^\prime}$ and $\alpha_{ji}$ in Eq.~(\ref{Eq:En9}) are defined as $J_1$ and $kd$ for a nearest-neighbor layer and by $J_2$ and $2kd$ for a next-nearest-neighbor layer, respectively, $k$ is the magnitude of the helix wavevector along the $z$~axis and $d$ is the distance between layers as shown in Fig.~\ref{Fig:J0_Jz1_Jz2_model_helix}.  The prefactors of two in the last two terms occur because each layer has two nearest-layer neighbors and two next-nearest-layer neighbors.  The turn angle $kd$ between adjacent FM-aligned layers in the helix in zero appied field is given in terms of $J_1$ and $J_2$ by~\cite{Johnston2015}
\be
\cos(kd) = -\frac{J_1}{4J_2},
\label{Eq:coskd}
\ee
which we utilize in subsequent calculations in this paper.

This paper is particularly concerned with spin-flop phases that can arise from an external field~$H_x$ that is perpendicular to the helix $z$~axis for which the moments are not confined to the $xy$ plane but also have $z$ components.  In that case, we still assume that all moments in a layer perpendicular to the helix $z$ axis are FM aligned, but that the $z$~component can vary from layer to layer.  Therefore for the spin-flop phase, the exchange energy per spin in Eq.~(\ref{Eq:EnJ0J1J2}) is generalized to read
\bea
E_{{\rm exch}n} &=& \frac{S^2}{2}\big[J_0 + J_{1}\hat{\mu}_n\cdot\left(\hat{\mu}_{n+1}  + \hat{\mu}_{n-1}\right)\label{Eq:ExchnJ0J1J2}\hspace{0.2in}\\
&&\hspace{0.42in} +\ J_2\hat{\mu}_n\cdot\left(\hat{\mu}_{n+2}  + \hat{\mu}_{n-2}\right)\big].\nonumber
\eea
This equation  reduces to Eq.~(\ref{Eq:EnJ0J1J2}) if the $z$~components of the $\hat{\mu}_i$ are zero and the turn angle between the moments in adjacent layers is~$kd$ as in the helix in Fig.~\ref{Fig:J0_Jz1_Jz2_model_helix} when the external applied field is \mbox{$H_x = 0$}. 

It is convenient to normalize all exchange constants by $J_2$ because $J_2>0$ for a helix \cite{Johnston2015}.  Defining the dimensionless ratios
\be
J_{02} \equiv \frac{J_0}{J_2}, \quad J_{12} \equiv \frac{J_1}{J_2}, \quad J_{22} \equiv \frac{J_2}{J_2} \equiv 1,
\ee
Eq.~(\ref{Eq:ExchnJ0J1J2}) becomes
\bea
E_{{\rm exch}n} &=& \frac{S^2J_2}{2}\big[J_{02} + J_{12}\hat{\mu}_n\cdot\left(\hat{\mu}_{n+1}  +\hat{\mu}_{n-1}\right)\label{Eq:ExchnJ0J1J2B}\hspace{0.2in}\\
&&\hspace{0.61in} +\ \hat{\mu}_n\cdot\left(\hat{\mu}_{n+2}  + \hat{\mu}_{n-2}\right)\big].\nonumber
\eea

Then normalizing all energies by $S^2J_2$~\cite{Johnston2017c}, Eq.~(\ref{Eq:EnTotal}) for the energy per spin now reads
\bea
\frac{E_n}{S^2J_2} &=&  \frac{1}{2}\big[J_{02} + J_{12}\hat{\mu}_n\cdot\left(\hat{\mu}_{n+1}  +\hat{\mu}_{n-1}\right)\label{Eq:ExchnJ0J1J2B}\hspace{0.2in}\label{Eq:EnTotal2}\\
&&\hspace{0.39in} +\ \hat{\mu}_n\cdot\left(\hat{\mu}_{n+2}  + \hat{\mu}_{n-2}\right)\big]\nonumber\\
&& -\ \frac{3S}{2(S+1)} \frac{g\mu_{\rm B}H_{\rm A1}}{S^2J_2}(1 - \bar{\mu}_{nz}^2) \nonumber\\
&& -\ \bar{\mu}_{nx} S\frac{g\mu_{\rm B}H_x}{S^2J_2}. \nonumber
\eea

Dimensionless reduced magnetic fields are defined as
\bse
\label{Eqs:hxhA1}
\bea
h_x^* &=& \frac{g\mu_{\rm B}H_x}{S^2J_2},\\
h_x^{**} &=& \frac{Sg\mu_{\rm B}H_x}{S^2J_2} = \frac{g\mu_{\rm B}H_x}{SJ_2} = Sh_x^*,\label{Eq:hx**Def}\\
h_{\rm A1}^* &=& \frac{g\mu_{\rm B}H_{\rm A1}}{S^2J_2},\\
h_{\rm A1}^{**} &=&  \frac{3Sg\mu_{\rm B}H_{\rm A1}}{2(S+1)S^2J_2}= \frac{3S}{2(S+1)}h_{\rm A1}^*,\label{Eq:hA1**Def}\\
h_{\rm c}^* &=& \frac{g\mu_{\rm B}H_{\rm c}}{S^2J_2},\\
h_{\rm c}^{**} &=& \frac{Sg\mu_{\rm B}H_{\rm c}}{S^2J_2} = \frac{g\mu_{\rm B}H_{\rm c}}{SJ_2} = Sh_{\rm c}^*,
\eea
\ese
where the last two expressions are for the reduced critical field~$h_{\rm c}$ discussed in the following Sec.~\ref{Sec:SFPhase}.  Using Eqs.~(\ref{Eqs:hxhA1}), the normalized energy in Eq.~(\ref{Eq:ExchnJ0J1J2B}) becomes
\bse
\label{Eqs:EnEave}
\bea
\frac{E_n}{S^2J_2} &=&  \frac{1}{2}\big[J_{02} + J_{12}\hat{\mu}_n\cdot\left(\hat{\mu}_{n+1}  +\hat{\mu}_{n-1}\right)\label{Eq:ExchnJ0J1J2C}\hspace{0.2in}\label{Eq:EnTotal2}\\
&&\hspace{0.6in} +\ \hat{\mu}_n\cdot\left(\hat{\mu}_{n+2}  + \hat{\mu}_{n-2}\right)\big]\nonumber\\
&& \hspace{-0.1in} -\ \big[h_{\rm A1}^{**}\left(1 - \bar{\mu}_{nz}^2 \right) + \bar{\mu}_{nx}h_x^{**}\big].\nonumber
\eea
Thus a nonzero out-of-plane component $\bar{\mu}_{nz}$ of a moment unit vector $\hat{\mu}_n$ in Eq.~(\ref{Eq:hatmun}) increases the energy of that moment, as expected for XY anisotropy.  However, we find below that the negative contribution of the $h_x^{**}$ term can offset the former positive contribution, leading to a net decrease in the normalized average energy per moment
\be
\frac{E_{\rm ave}}{S^2J_2} = \frac{1}{n_\lambda}\sum_{n=1}^{n_\lambda}\frac{E_n}{S^2J_2},
\label{Eq:EaveDef}
\ee
\ese
where $n_\lambda$ is the integer number of moment layers per commensurate wavelength that is assumed for the in-plane helix.

In order to compare the value of $E_{\rm ave}/(S^2J_2)$ with that calculated at $T=0$ for an in-plane helix/fan for the same $h_x^{**}$~\cite{Johnston2017c}, in Eq.~(\ref{Eq:EnTotal2}) we set
\bse
\label{Eqs:J12kd}
\be
J_{12} = -4\cos(kd)
\ee
according to Eq.~(\ref{Eq:coskd}), where
\be
kd = 2\pi m/n_\lambda
\ee
\ese
is the turn angle in Fig.~\ref{Fig:J0_Jz1_Jz2_model_helix} between adjacent layers of a helix in zero applied field with integer \mbox{$m < n_\lambda$,} and is assumed to be independent of both the applied and anisotropy fields.  For this comparison, we also set
\be
J_{02}=0.
\ee

\section{\label{Sec:SFPhase} Results: The Spin-Flop Phase}

Values of the average energy per spin and the average $x$-axis magnetic moment per spin versus the reduced field $h_x^{**}$ when the moments in a zero-field helix and high-field fan are confined to the $xy$~plane were calculated for $T=0$ in Ref.~\cite{Johnston2017c}.  Here we calculate these $T=0$ properties for the spin-flop (SF) phase where the moments flop out of the $xy$~plane due to a nonzero $h_x^{**}$. A comparison of the average energy per spin in the helix and SF phases versus $h_x^{**}$ and~$h_{\rm A1}^{**}$ will be needed for the construction of the $T=0$ phase diagrams in the $h_x^{**}$-$h_{\rm A1}^{**}$ plane in Sec.~\ref{Sec:Phase Diagrams}.

In the absence of an anisotropy field, in zero applied field a hodograph of the moments in a helix is a circle in the $xy$ plane as shown in Fig.~\ref{Fig:J0_Jz1_Jz2_model_helix}.  For an infinitesimal $h_x^{**}$, the moments flop by 90$^\circ$ into the $yz$~plane, thus forming a circular hodograph in the $yz$ plane.  However, in the presence of a finite XY anisotropy field~$h_{\rm A1}^{**}$, we assume that the latter circle is flattened into an ellipse in the $yz$~plane where the semimajor axis $a$ of the ellipse is along the $y$~axis and the semiminor axis $b$ is along the $z$~axis.  Due to the fact that we only consider $T=0$, the moment magnitude $\mu$ is fixed at the value given in Eq.~(\ref{Eq:muDef}).  Hence a hodograph of the moment unit vectors $\hat{\mu}$ in the presence of a nonzero $h_x^{**}$ is a spherical ellipse of radius unity, which is the projection of a two-dimensional ellipse in the $yz$ plane onto a sphere of radius unity. The magnitude~$\mu$ of the magnetic moments is taken into account in the reduced fields $h_x^{**}$ and $h_{\rm A1}^{**}$ in Eqs.~(\ref{Eqs:hxhA1}) above.

\begin{figure}
\includegraphics [width=2.55in]{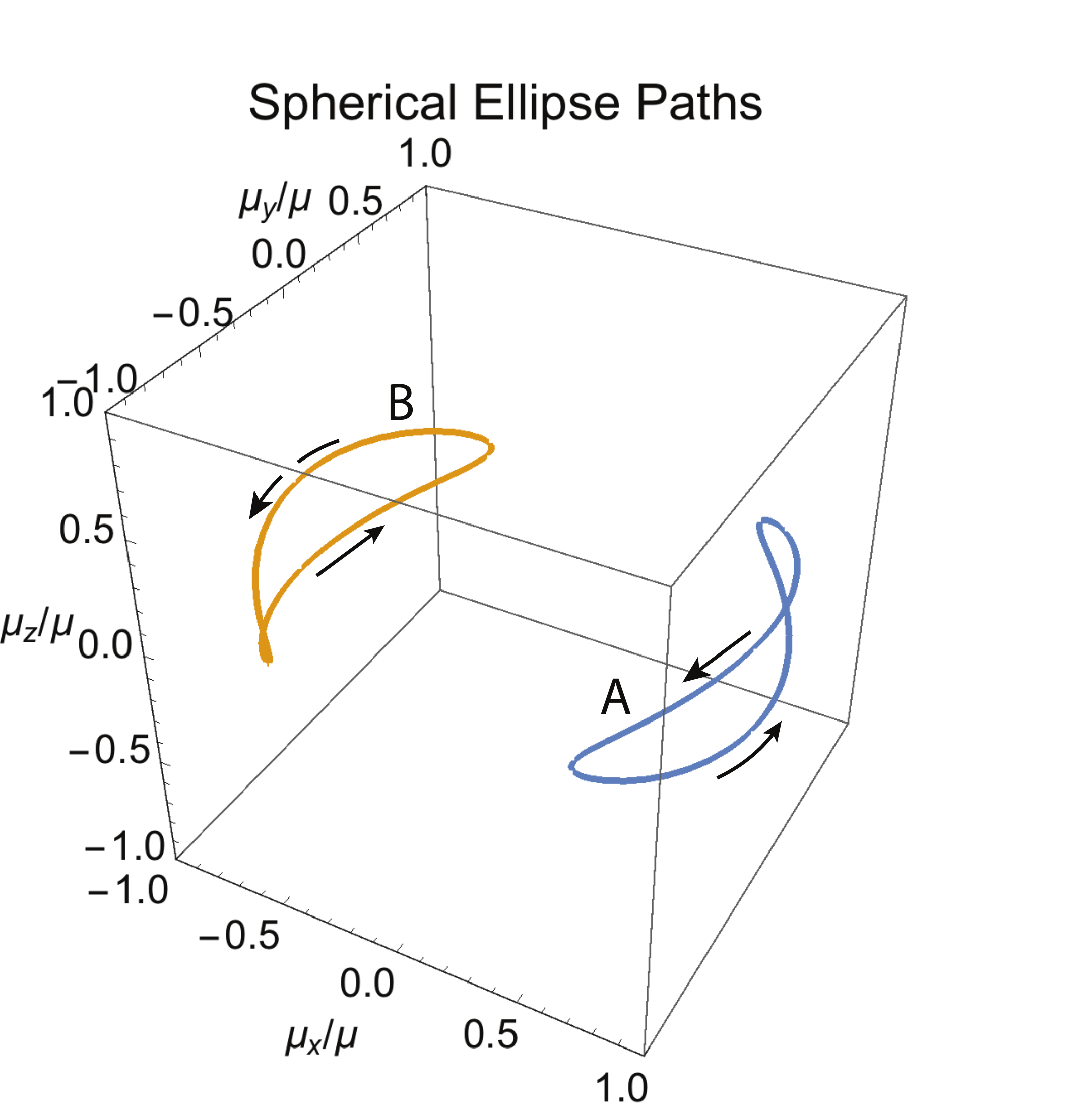}
\caption {Spherical ellipse paths (hodographs) of the magnetic moment unit vectors $\vec{\mu}/\mu$ in sublattices A and B in the spin-flop (SF) phase according to Eqs.~(\ref{Eqs:AB paths}) with the parameter $c=-1$.  These paths are elliptical in the $yz$ plane with a constant radius of unity from the origin of the Cartesian coordinate system.  In this illustration, the semimajor and semiminor axes of the elliptic paths in the $yz$~plane are set to $a_{\rm A} = a_{\rm B} = 0.8,\ b_{\rm A} = b_{\rm B} = 0.2$, but the equalities $a_{\rm A} = a_{\rm B}$ and $b_{\rm A} = b_{\rm B}$ are generally not obtained for the SF phase from energy minimization even when $c=1$ and the spherical ellipses are both on the positive side of $x=0$ towards which the applied magnetic field {\bf H} points.}
\label{Fig:Spherical_Ellipse_Paths}
\end{figure}

In the spin-flop phase with finite $h_x^{**}$ and $h_{\rm A1}^{**}$, one expects at least for the case of AFM $J_{\rm 12}>0$ with the applied field in Eq.~(\ref{Eq:HvecDef}), that two spherical elliptic paths (hodographs) A and B traversed by the magnetic-moment unit vectors could occur in which the $x$ components have opposite signs in order to decrease the value of exchange interaction energy between spins in adjacent layers.  Then the reduced moments with even $n$ in sublattice A are described by
\bse
\label{Eqs:AB paths}
\bea
\hat{\mu}_{{\rm A}n} &=& \bar{\mu}_{{\rm A}nx}\,\hat{\bf i} +\bar{\mu}_{{\rm A}ny}\,\hat{\bf j} + \bar{\mu}_{{\rm A}nz}\,\hat{\bf k} ,\quad (n\ {\rm even})\hspace{0.2in}\\
\bar{\mu}_{{\rm A}ny} &=& a_{\rm A}\cos(nkd),\\
\bar{\mu}_{{\rm A}nz} &=& b_{\rm A}\sin(nkd),\\
\bar{\mu}_{{\rm A}nx} &=& \sqrt{1 - (\bar{\mu}_{{\rm A}ny}^2 + \bar{\mu}_{{\rm A}nz}^2)},\label{Eq:munxA}
\eea
and the moments in sublattice B with odd~$n$ are described by
\bea
\hat{\mu}_{{\rm A}n} &=& \bar{\mu}_{{\rm B}nx}\,\hat{\bf i} + \bar{\mu}_{{\rm B}ny}\,\hat{\bf j} + \bar{\mu}_{{\rm b}nz}\,\hat{\bf k} ,\quad (n\ {\rm odd})\hspace{0.2in}\\
\bar{\mu}_{{\rm B}ny} &=& a_{\rm B}\cos(nkd),\\
\bar{\mu}_{{\rm B}nz} &=& b_{\rm B}\sin(nkd),\\
\bar{\mu}_{{\rm B}nx} &=& c\sqrt{1 - (\bar{\mu}_{{\rm B}ny}^2 + \bar{\mu}_{{\rm B}nz}^2)},\label{Eq:munxB}
\eea
\ese
where $c=\pm1$, $n = 1,\ 2,\ \ldots, n_\lambda$, and for each~$n$ within each sublattice Eq.~(\ref{Eq:munDotmun}) is satisfied.  The moments are distributed in equal numbers between sublattices A and~B, labeled by consecutive odd and even integers~$n$, respectively, so the total number of moments $n_\lambda$ per wavelength $\lambda = n_\lambda d$ along the $z$~axis is even.  An illustration of the spherical ellipse paths (hodographs) of sublattices A and~B described by Eqs.~(\ref{Eqs:AB paths}) is shown in Fig.~\ref{Fig:Spherical_Ellipse_Paths} for $c=-1,\ a_{\rm A} = a_{\rm B} = 0.8$ and $b_{\rm A} = b_{\rm B} = 0.2$.  The value $c=-1$ corresponds to two spherical-elliptic paths on opposite sides of $\bar{\mu}_x=0$ for sublattices A and~B as shown in the figure.  This may be expected at small $h_x^{**}$ for AFM $J_1>0$, whereas when $c=1$ the paths are on the same sode of the positive $\bar{\mu}_x$~axis towards which the applied magnetic field {\bf H} points, as expected for all moments for large $h_x^{**}$ with either AFM or FM~$J_1$.

The spherical-ellipse parameters $c_,\ a_{\rm A},\ b_{\rm A},\ a_{\rm B},\ b_{\rm B}$ are all determined at the same time by minimizing the normalized average energy per spin $E_{\rm ave}/(S^2J_2)$ in Eq.~(\ref{Eq:EaveDef}) with respect to these parameters in Eqs.~(\ref{Eqs:AB paths}) when inserted into Eq.~(\ref{Eq:EnTotal2}) for fixed values of $h_{\rm A1}^{**}$ and $h_x^{**}$.  If the obtained values satisfy $c=-1$ or $c=1$ with $a_{\rm A}\neq a_{\rm B},\ b_{\rm A}\neq b_{\rm B}$, then there are two spherical ellipses, one on each side of $\bar{\mu}_x=0$ if $c=-1$ and both on the $\bar{\mu}_x>0$ side if $c=1$.  On the other hand, if $b_{\rm A}$ and $b_{\rm B}$ satisfy $b_{\rm A}=b_{\rm B}=0$ (no $z$-axis component to the moments), either one ($a_{\rm A}= a_{\rm B}$) or two ($a_{\rm A}\neq a_{\rm B}$) $xy$~fan phases are found.  Finally, if $a_{\rm A} = a_{\rm B} = b_{\rm A}=b_{\rm B}=0$, the moments all point in the direction of the applied field in the $+x$~direction and the system is in the PM state.

Once the spherical-ellipse parameters are determined, the average value of $x$~component of the magnetic moment unit vector in the direction of the applied field for the given values of $h_x^{**}$ and $h_{\rm A1}^{**}$ is obtained from
\be
\bar{\mu}_{x{\rm ave}} \equiv \frac{{\mu}_{x{\rm ave}}}{\mu} = \frac{1}{n_\lambda}\sum_{n=1}^{n_\lambda}\bar{\mu}_{nx}
\label{Eq:barmuxave}
\ee
using Eqs.~(\ref{Eq:munxA}) and~(\ref{Eq:munxB}).

\begin{figure*}
\includegraphics [width=7in]{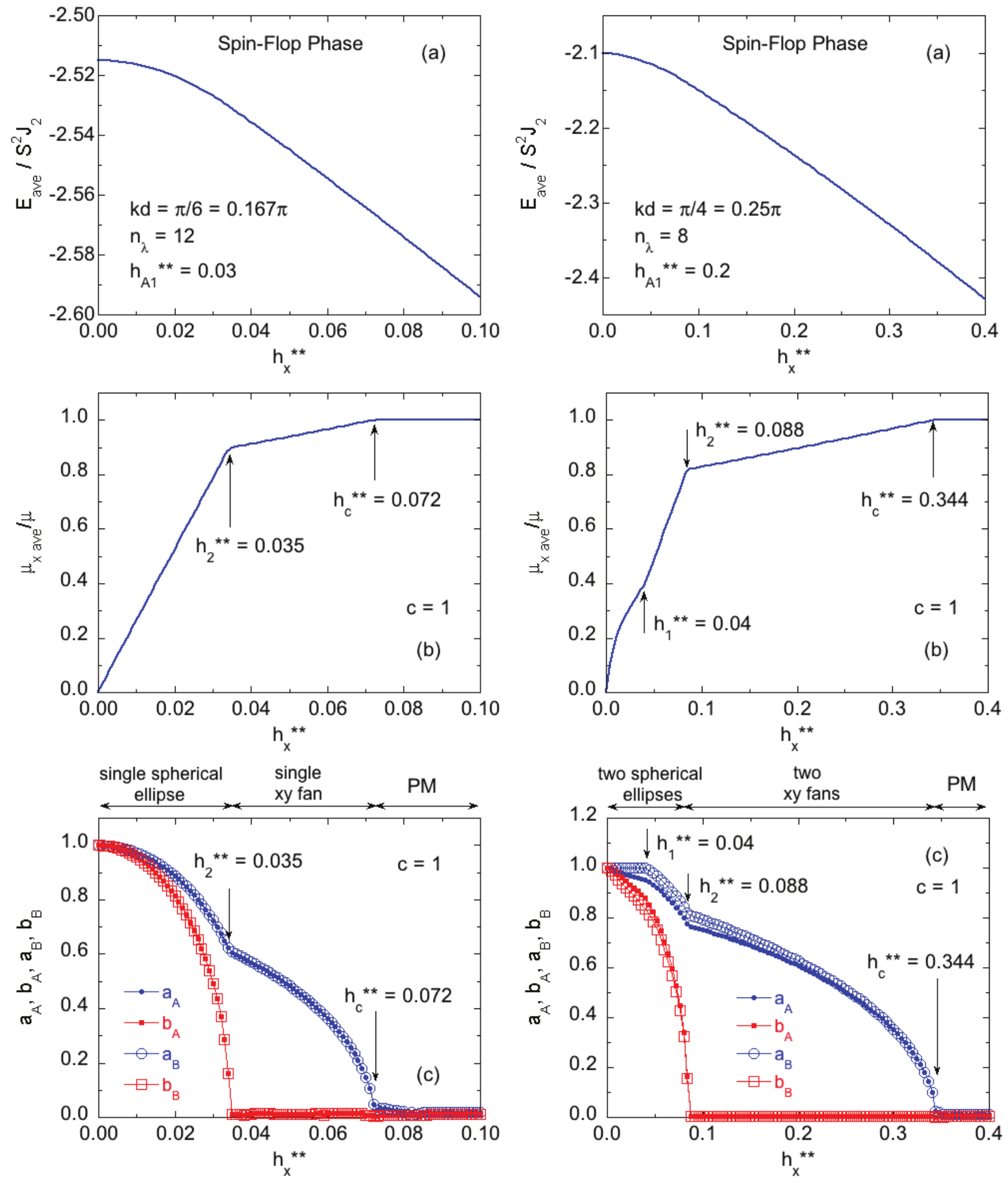}
\caption {(a) Evolution in the average energy ${\rm E_{ave}}$ normalized by $S^2J_2$, (b)~average $x$~component of the magnetic moment per spin $\mu_{\rm x\ ave}$ where we find $c=1$, and (c)~the spherical ellipse axes $a_{\rm A},\ b_{\rm A},\ a_{\rm B},\ b_{\rm B}$ and the corresponding phases labeled above the subfigure for interlayer turn angles $kd = \pi/6$ (left panels) and $kd = \pi/4$ (right panels).}
\label{Fig:kdPiOn6PiOn4}
\end{figure*}

\begin{figure*}
\includegraphics [width=7in]{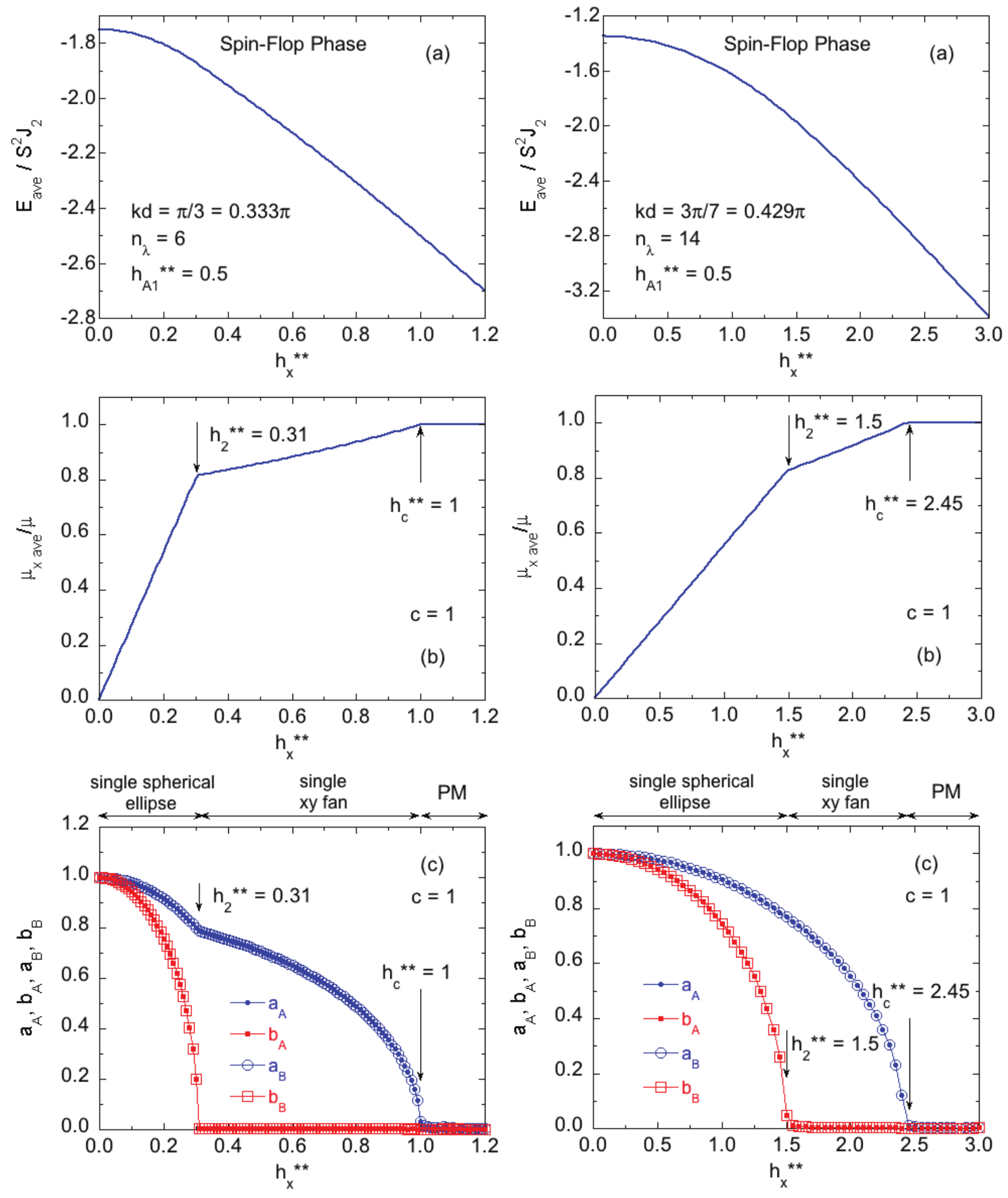}
\caption {Same as Fig.~\ref{Fig:kdPiOn6PiOn4} except for interlayer turn angles $kd = \pi/3$ (left panels) and $kd = 3\pi/7$ (right panels).}
\label{Fig:kdPiOn33PiOn7}
\end{figure*}

\begin{figure*}
\includegraphics [width=7in]{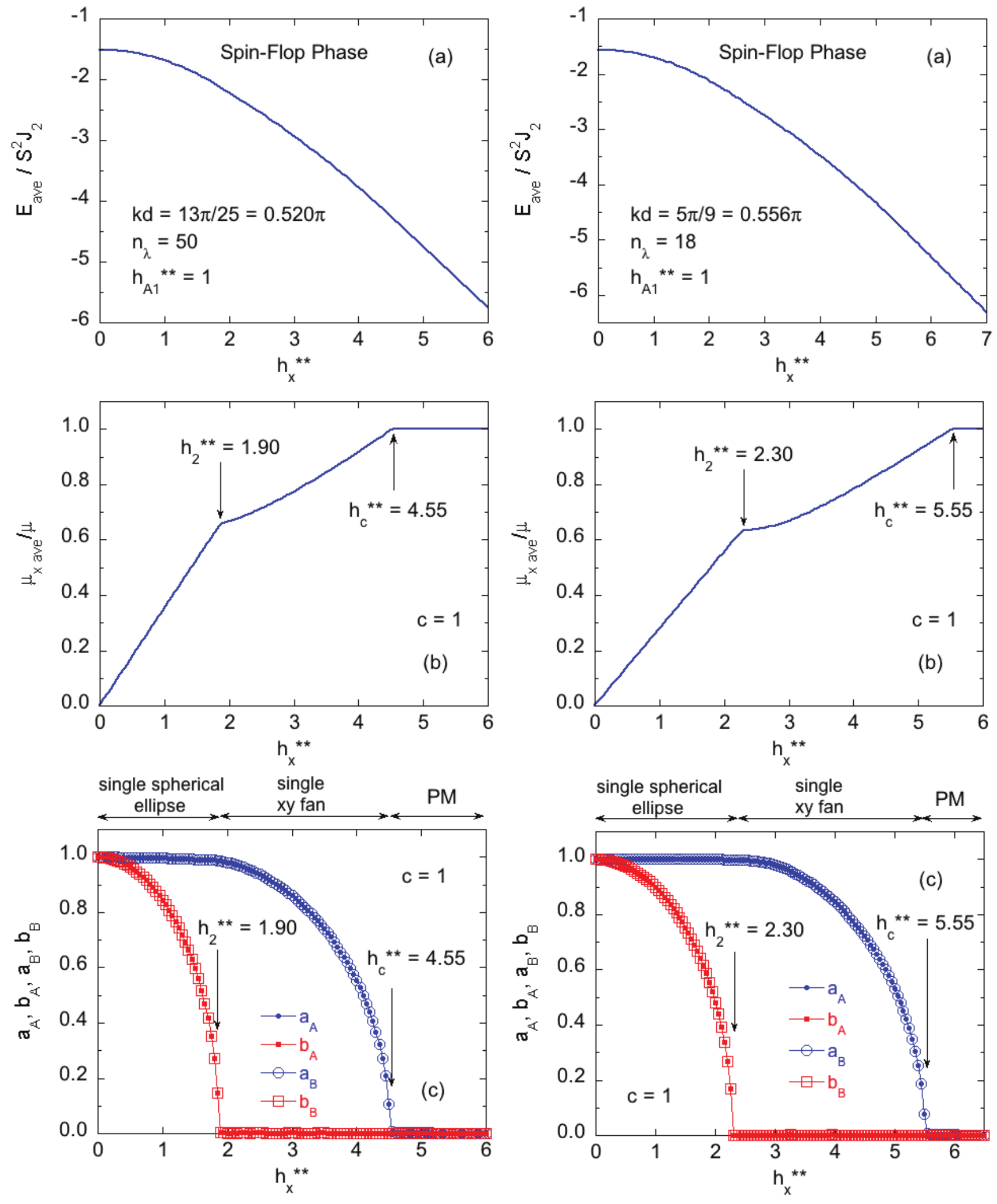}
\caption {Same as Fig.~\ref{Fig:kdPiOn6PiOn4} except for interlayer turn angles $kd = 13\pi/25$ (left panels) and $kd = 5\pi/9$ (right panels).}
\label{Fig:kd13PiOn255PiOn9}
\end{figure*}

\begin{figure*}
\includegraphics [width=7in]{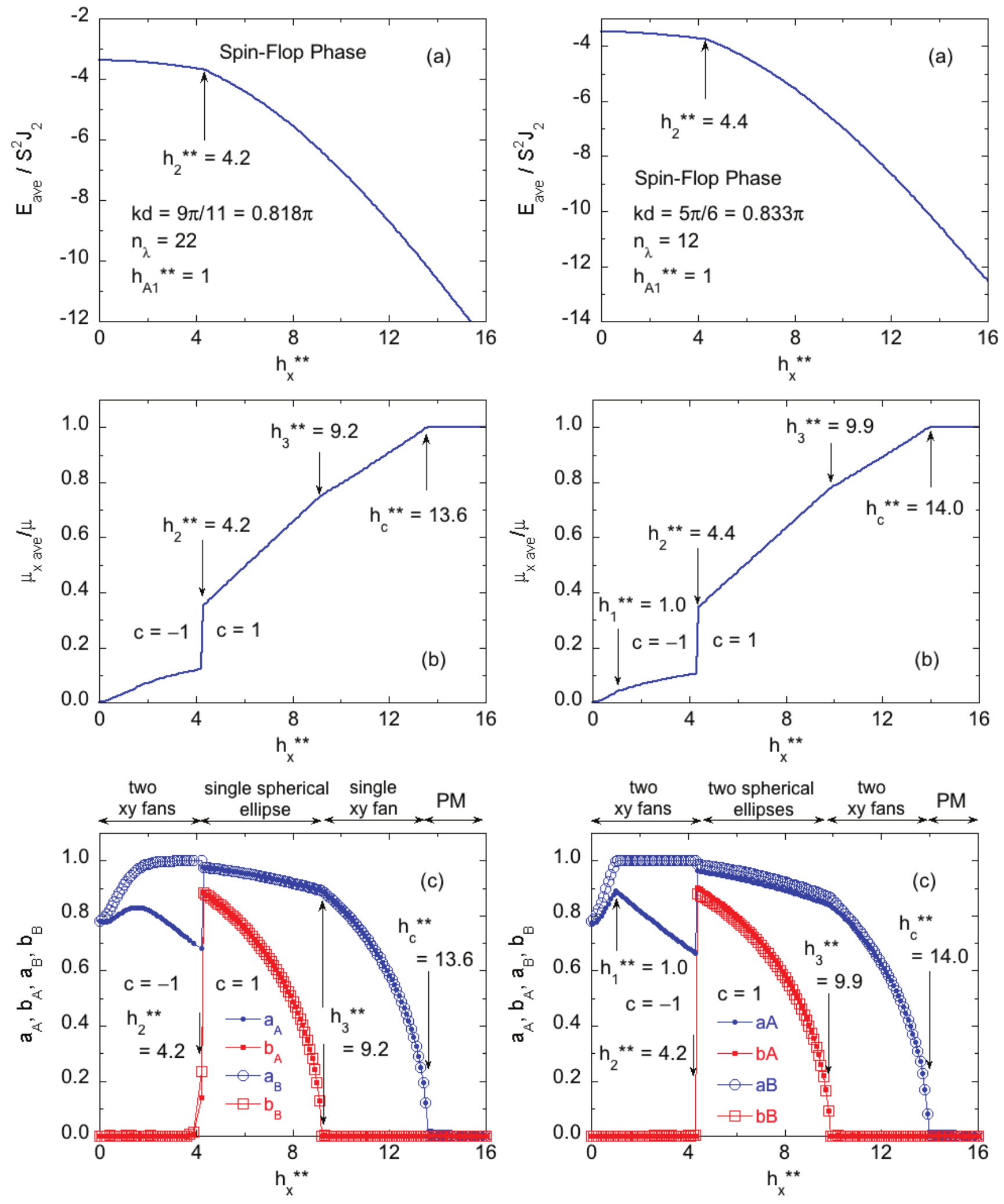}
\caption {Same as Fig.~\ref{Fig:kdPiOn6PiOn4} except for interlayer turn angles $kd = 9\pi/11$ (left panels) and $kd = 5\pi/6$ (right panels).}
\label{Fig:kd9PiOn115PiOn6}
\end{figure*}

The fitted values of $E_{\rm ave}/(S^2J_2)$, $\bar{\mu}_{x{\rm ave}}$, and of $c_,\ a_{\rm A},\ b_{\rm A}\ a_{\rm B},\ b_{\rm B}$ are shown for representative values $kd = \pi/6$ and~$\pi/4$; $\pi/3$ and $3\pi/7$; $13\pi/25$ and $5\pi/9$; and $9\pi/11$ and~$5\pi/6$; in Figs.~\ref{Fig:kdPiOn6PiOn4}(a)--\ref{Fig:kdPiOn6PiOn4}(c) to Figs.~\ref{Fig:kd9PiOn115PiOn6}(a)--\ref{Fig:kd9PiOn115PiOn6}(c), respectively.  One sees a variety of possible SF phases for different values of $h_{\rm A1}^{**}$ and  of~$h_x^{**}$, including a single spherical ellipse, a single $xy$ fan, two spherical ellipses, two $xy$ fans, and at high fields, the PM phase in which all moments are FM-aligned in the direction~$\hat{\bf i}$ of the applied field.  There is no clear monotonic dependence versus~$kd$ in the order in which the first four phases occur.  A nonmonotonic behavior versus $kd$ was previously found in the range $4\pi/9 \leq kd < \pi$ for the phases occuring at $T=0$ versus applied $x$-axis field for the $xy$~helix and $xy$~fan phases when the moments are confined to the $xy$~plane \cite{Johnston2017c}.  The stable phases for $0<kd<\pi/2$ with FM (negative) $J_{12}$ all have $c=1$ for all $h_x^{**}$ as anticipated, whereas two of the stable phases for $\pi/2<kd<\pi$ with AFM (positive) $J_{12}$ have $c=-1$ at low fields, as also anticipated, and $c=1$ at high fields.

First-order transitions versus $h_x^{**}$ occur when $c$ discontinuously changes with increasing $h_x^{**}$ from $-1$ to~1 in Fig.~\ref{Fig:kd9PiOn115PiOn6} for $kd=9\pi/11$ and $kd = 5\pi/6$.  The first-order nature of the transitions is also revealed in the $h_x^{**}$ dependences of $E_{\rm ave}$, $\bar{\mu}_{x\,{\rm ave}}$ and the other four spherical ellipse parameters.  The transitions versus $h_x^{**}$ for the other six $kd$ values in Figs.~\ref{Fig:kdPiOn6PiOn4} to \ref{Fig:kd13PiOn255PiOn9} are seen to be second order.  When $kd$ increases from $9\pi/11$ to $5\pi/6$, both with $h_{\rm A1}^{**}=1$ in Fig.~\ref{Fig:kd9PiOn115PiOn6}, a new second-order transition at $h_1^{**} = 1.0$ occurs for $kd = 5\pi/6$, whereas for $kd=9\pi/11$ the transition is instead a smooth crossover.

The reduced critical field $h_{\rm c}^{**}$ versus~$kd$ is the value at which the system becomes PM with increasing~$h_x^{**}$.  These second-order transition fields are listed for each of the eight $kd$ values and the specified values of $h_{\rm A1}^{**}$ in Figs.~\ref{Fig:kdPiOn6PiOn4} to \ref{Fig:kd9PiOn115PiOn6}.  We find that $h_{\rm c}^{**}$ only depends only on $kd$ (not on $h_{\rm A1}^{**}$), when $n_\lambda$ is even as assumed in this paper.  For $h_x^{**}\to h_{\rm c}^{**-}$, the stable phase for all values of $kd$ is a single fan in the $xy$ plane, which was studied in detail in Ref.~\cite{Johnston2017c}.  The approximate values of $h_{\rm c}^{**}$ versus~$kd$ listed in Figs.~\ref{Fig:kdPiOn6PiOn4} to \ref{Fig:kd9PiOn115PiOn6} are in agreement with the respective exact values given for the $xy$ fan by~\cite{Johnston2017c}
\bse
\bea
h_{\rm c}^{**} &=& 16\sin^4\left(\frac{kd}{2}\right)\hspace{0.4in} (0\leq kd \leq \pi/2),\hspace{0.3in},\\
h_{\rm c}^{**} &=& 16\cos^4\left(\frac{\pi - kd}{2}\right)\quad (\pi/2\leq kd \leq \pi).\label{Eq:hc**kd>pi/2}
\eea
\ese

\section{\label{Sec:Phase Diagrams} Phase Diagrams in the $\bf h_x^{**}$-$\bf h_{\rm A1}^{**}$ Plane for Representative $\bf kd$ Values}

As discussed above, the phases that can occur within MFT are the $xy$~helix phase with moments aligned in the $xy$~plane ($xy$~helix/fan), the spin-flop (SF) phase with moments that have three-dimensional components ($xyz$~spin flop), the $xy$~fan phase with moments oriented within the $xy$~plane ($xy$ fan) and the paramagnetic (PM) phase where the moments are ferromagnetically-aligned in the direction of the $x$-axis reduced field $h_x^{**}$.

The phase boundary between the $xy$~helix phase and the $xy$~fan phase of the helix when it occurs was determined previously in Ref.~\cite{Johnston2017c}, where the energies of the $xy$~helix and higher-field $xy$~fan phases were  determined versus~$h_x^{**}$ in Ref.~\cite{Johnston2017c}.  However, here one needs to determine the influence of $h_{\rm A1}^{**}$ on those energies.  Since these moments are confined to the $xy$~plane, the reduced energy of moment layer~$n$ for the $xy$~helix and associated high-field $xy$~fan phases is given by Eq.~(\ref{Eq:EnTotal2}) as 
\bse
\bea
\frac{E_n^{\rm helix/fan}}{S^2J_2} &=& \frac{1}{2}\big[J_{12}\left(\hat{\mu}_n\cdot\hat{\mu}_{n+1}  + \hat{\mu}_n\cdot\hat{\mu}_{n-1}\right) \label{Eq:EnHelixFan}\\
&& \hspace{0.3in} +\ \big(\hat{\mu}_n\cdot\hat{\mu}_{n+2} + \hat{\mu}_n\cdot\hat{\mu}_{n-2} \big)\big]\nonumber\\&& \hspace{-0.1in} -\ \big(h_{\rm A1}^{**} + \bar{\mu}_{nx}h_x^{**}\big)\nonumber\\
&=& \frac{E_n^{\rm helix/fan}}{S^2J_2}(h_x^{**},h_{\rm A1}^{**} = 0) -h_{\rm A1}^{**},\label{Eq:Eavehelixfan}\hspace{0.3in}
\eea
\ese
where the first term on the right-hand side of the bottom equality was calculated for a variety of turn angles~$kd$ in Ref.~\cite{Johnston2017c}.

One anticipates that when $h_{\rm A1}^{**} = 0$, in order for the system to minimize its energy an infinitesimal $h_x^{**}$ causes the $xy$~helix to immediately spin-flop to a perpendicular orientation in the $yz$~plane.  With further increases in $h_x^{**}$, the moments all tilt by the same angle towards the $x$~axis as shown in Fig.~\ref{Fig:High_Perp_Field_Struct} where the $z$~axis in that figure is replaced by the $x$~axis here.  When $h_{\rm A1}^{**}$ increases to a finite value, one expects a finite field to be required to cause the moments to flop out of the $xy$~plane to enter the SF phase.  However, if $h_{\rm A1}^{**}$ is sufficiently large, this $xy$~helix to $xyz$~spin-flop transition is expected to be replaced by the previously-studied $xy$~helix to $xy$~fan phase transition.  These expectations are borne out by the phase diagrams shown in Fig.~\ref{Fig:helix_SF_phase_diagrams} below.

The reduced phase transition field $h_x^{**}$ between the $xy$~helix phase and the $xyz$~spin-flop phase for a given value of reduced XY anisotropy field $h_{\rm A1}^{**}$ was determined by the crossover in average energy between these two phases, where at low fields the $xy$~helix phase has the lower energy and at higher fields the $xyz$~spin-flop phase energy is lower.  This is a first-order transition.  The transition between the $xy$~helix phase and the high-field $xy$~fan phase can be first-order,  second-order, or a smooth crossover \cite{Johnston2017c}.  The phase transition field between the $xyz$~spin-flop phase and the PM phase or between the $xy$~fan phase and the PM phase are determined by the criterion that the $x$~component of the calculated average moment unit vector per spin $\bar{\mu}_{x\,{\rm ave}} \equiv {\mu}_{x\,{\rm ave}}/\mu$ becomes equal to unity with increasing $h_x^{**}$.  This is a continuous (second-order) transition.

\begin{figure*}
\includegraphics [width=7in]{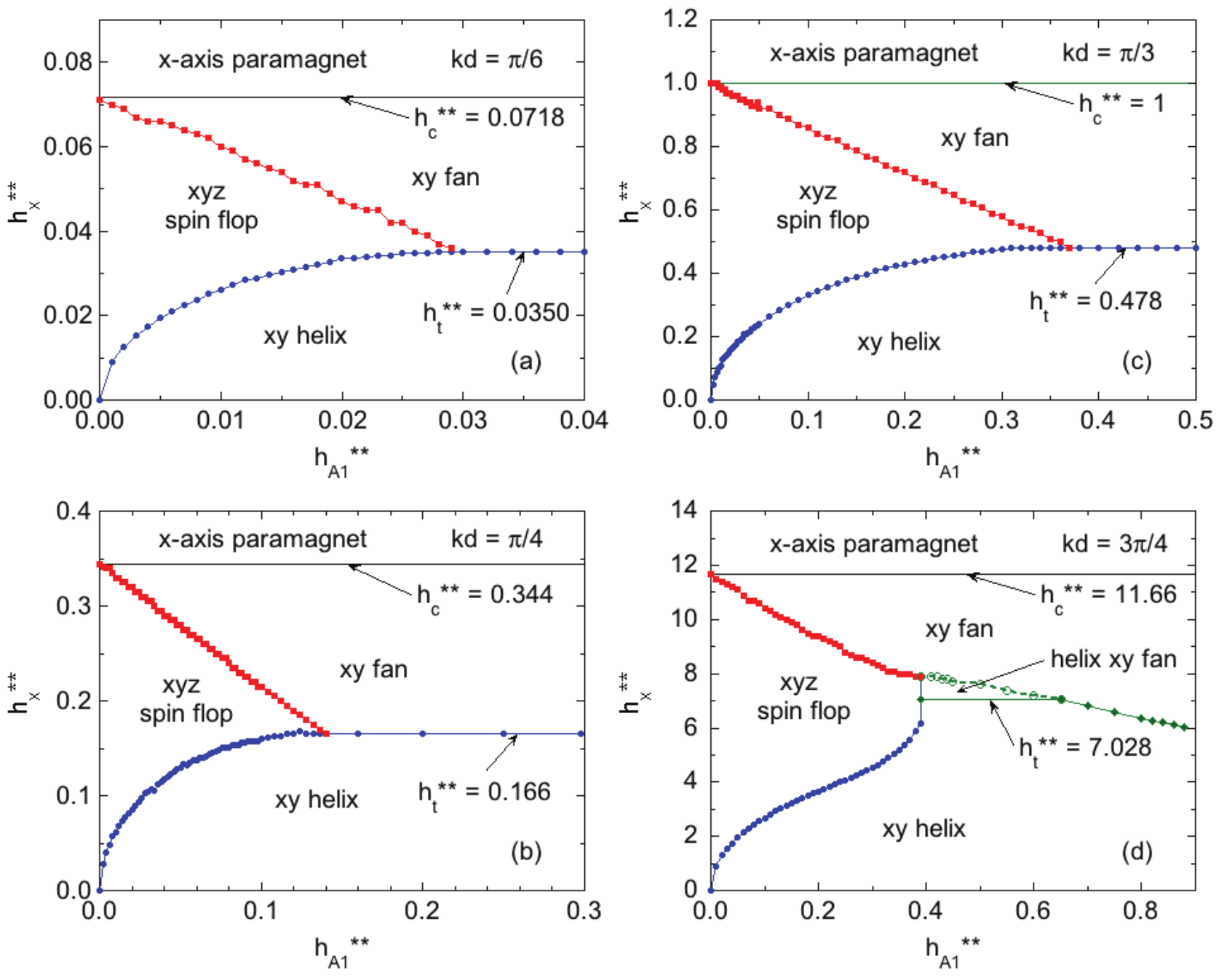} 
\caption {Phase diagrams in the $h_x^{**}-h_{\rm A1}^{**}$ plane at temperature $T=0$ for interlayer turn angles (a)~$kd = \pi/6$, (b)~$kd = \pi/4$, (c)~$kd = \pi/3$, and (d)~$kd = 3\pi/4$.  For these values of $kd$, the phase transitions from the $xy$~helix to the $xyz$ spin-flop phase and from the $xy$~helix to the $xy$~fan at a reduced field $h_{\rm t}^{**}$ are first order, whereas the transitions from the $xyz$~spin-flop phase to the $xy$~fan phase and from the $xy$~fan phase to the paramagnetic phase at reduced field $h_{\rm c}^{**}$ are second order.}
\label{Fig:helix_SF_phase_diagrams}
\end{figure*}

The phase diagrams in the $h_x^{**}-h_{\rm A1}^{**}$ plane at $T=0$ calculated for the four turn angles $kd = \pi/6,\ \pi/4,\ \pi/3$ and~$3\pi/4$ are shown in Figs.~\ref{Fig:helix_SF_phase_diagrams}(a), \ref{Fig:helix_SF_phase_diagrams}(b), \ref{Fig:helix_SF_phase_diagrams}(c), and \ref{Fig:helix_SF_phase_diagrams}(d), respectively.  The first three turn angles correspond to FM nearest-layer couplings~$J_1<0$ whereas the fourth one is for an AFM $J_1>0$.  One sees that the phase diagrams follow the above expectations.  The first three phase diagrams with FM $kd<\pi/2$ have common forms, where approximately the same phase diagram is obtained but with a rescaling of the $h_x^{**}$ and $h_{\rm A1}^{**}$ axes.  In all three phase diagrams the phase transition line between the $xyz$ spin flop and the $xy$~fan phases is linear or nearly so.  Another interesting feature is that all three phase diagrams show a horizontal first-order $xy$~helix to $xy$~fan phase boundary at large $h_{\rm A1}^{**}$ values.  This occurs at the respective first-order transition fields $h_{\rm t}^{**}$ between these two $xy$ phases reported previously in Ref.~\cite{Johnston2017c}.  These three phase diagrams are similar in form to the $T=0$ phase diagram in Fig.~4 of Ref.~\cite{Nagamiya1962} for small values of~$kd$.

The phase diagram for $kd = 3\pi/4$ in Fig.~\ref{Fig:helix_SF_phase_diagrams}(d) for $kd > \pi/2$ corresponding to AFM~$J_1>0$ is different from Figs.~\ref{Fig:helix_SF_phase_diagrams}(a) to~\ref{Fig:helix_SF_phase_diagrams}(c) where the nearest-layer coupling is FM\@.  First, the phase transition line between the $xyz$ spin flop and the $xy$~fan phases with increasing $h_x^{**}$ at fixed $h_{\rm A1}$ in Fig.~\ref{Fig:helix_SF_phase_diagrams}(d) is not linear compared to the linear behavior in Figs.~\ref{Fig:helix_SF_phase_diagrams}(a) to~\ref{Fig:helix_SF_phase_diagrams}(c).  Second, the phase line in Fig.~\ref{Fig:helix_SF_phase_diagrams}(d) between the $xy$~helix and the $xyz$~spin-flop phase first exhibits negative curvature, but then shows an inflection point with positive curvature at larger values of $h_{\rm A1}^{**}$, whereas this phase line has uniformly negative curvature in Figs.~\ref{Fig:helix_SF_phase_diagrams}(a) to~\ref{Fig:helix_SF_phase_diagrams}(c).  Third, a second-order transition between the $xy$~helix and the $xy$~fan phase occurs on the right side of Figs.~\ref{Fig:helix_SF_phase_diagrams}(d), whereas in Figs.~\ref{Fig:helix_SF_phase_diagrams}(a) to~\ref{Fig:helix_SF_phase_diagrams}(c) the transition is first order.  Fourth, the phase transition line between the $xy$~helix and the $xy$~fan obtained as described above (green open circles and filled diamonds) has a negative slope, compared to the zero slope for the first three phase diagrams.  Finally, the negative-slope second-order phase boundary between the $xy$~helix and the $xy$~fan phases in the region $h_{\rm A1}^{**} \approx 0.4$ to 0.65 found by minimizing the energy between the helix and $xy$~fan phases is preempted by a horizontal first-order transition line found from energy minimization between the helix and fan phases in Ref.~\cite{Johnston2017c} for the case where the moments were constrained to lie in the $xy$~plane.  This is not seen in the first three phase diagrams.

We emphasize that the transitions versus $h_x^{**}$ at fixed $h_{\rm A1}^{**}$ for the SF phase shown in Figs.~\ref{Fig:kdPiOn6PiOn4} to~\ref{Fig:kd9PiOn115PiOn6} for particular values of $kd$ are only observed in a real helical Heisenberg AFM compound  if the SF phase has a lower energy than each of the $xy$~helix and $xy$~fan phases for the particular values of $kd$, $h_{\rm A1}^{**}$ and range of $h_x^{**}$ that are associated with the compound.  Indeed, we show that for the model helical Heisenberg antiferromagnet \ecp\ discussed in Sec.~\ref{Sec:EuCo2P2} below, the values of $kd$ and $h_{\rm A1}^{**}$ do not allow the SF phase to have a lower energy than the $xy$~helix or $xy$~fan phases for any value of $h_x^{**}$. Hence only the $xy$~helix,  $xy$~fan, and PM phases occur with increasing~$h_x^{**}$.

\section{\label{Sec:ExptThy} Comparison of the Theory with Experiment}

\subsection{\label{Sec:hA1**hx**Convert} Expressing $\bf h_{\rm A1}^{**}$ and~$\bf h_x^{**}$ in terms of experimental values of $\bf h_{\rm A1}$ and~$\bf h_x$}

In order to compare experimental magnetic data for helical Heisenberg antiferromagnets with the above theory, one needs to determine which region of the phase diagram ($xy$~helix phase, $xy$~fan phase, $xyz$~SF phase, or PM phase) a material lies for the material's values of $h_x^{**}$ and~$h_{\rm A1}^{**}$.  Then one can compare the experimental magnetization versus field data for the compound at low temperatures with the phase diagrams as in Fig.~\ref{Fig:helix_SF_phase_diagrams} to determine what phase transitions are predicted versus $x$-axis field for comparison with the experimental data.

To accomplish this comparison, one must first determine how the value of the reduced applied field~$h_{x}^{**}$ and anisotropy field $h_{\rm A1}^{**}$ in this paper are  expressed in terms of the reduced applied field $h_x$ and reduced anisotropy field $h_{\rm A1}$ defined in Ref.~\cite{Johnston2017b} that can be obtained from experimental magnetic susceptibility data (see following section).  From Ref.~\cite{Johnston2017b}, one has~
\bse
\be
h_{\rm A1} \equiv \frac{g\mu_{\rm B}H_{\rm A1}}{k_{\rm B}T_{{\rm N}J}},
\ee
where $k_{\rm B}$ is Boltzmann's constant and $T_{{\rm N}J}$ is the N\'eel temperature that would be obtained from Heisenberg exchange interactions alone with no anisotropy contributions.  A comparison of this definition with that for $h_{\rm A1}^{**}$ in Eq.~(\ref{Eq:hA1**Def}) gives the conversion
\be
h_{\rm A1}^{**} = \left[\frac{3}{2S(S+1)}\right]\left(\frac{k_{\rm B}T_{{\rm N}J}}{J_2}\right)h_{\rm A1}.
\label{Eq:hA1Conversion}
\ee
Similarly, a comparison of the definition~\cite{Johnston2017b}
\be
h_x \equiv   \frac{g\mu_{\rm B}H_x}{k_{\rm B}T_{{\rm N}J}}
\label{Eq:hxTOHx}
\ee
with that for $h_x^{**}$ in Eq.~(\ref{Eq:hx**Def}) yields
\be
h_x^{**} = \frac{1}{S}\left(\frac{k_{\rm B}T_{{\rm N}J}}{J_2}\right)h_x.
\label{Eq:hxConversion}
\ee
\ese
These conversions require the spin~$S$ to be known and also the material-specific ratio $k_{\rm B}T_{\rm {N}J}/J_2$ within the $J_0$-$J_1$-$J_2$ MFT model to be computed from magnetic susceptibility data for single crystals of the material. The latter calculation also yields $J_0$ and $J_1$ as discussed in the following section.

\subsection{\label{Sec:hA1etcExtraction} Extracting values of $\bf h_{\rm A1}$, $\bf T_{{\rm N}J}$, $\bf J_0$, $\bf J_1$, ad $\bf J_2$ from experimental magnetic susceptibility data within unified molecular-field theory}

The value of the XY anisotropy parameter $h_{\rm A1}$ is estimated from the anisotropy in the experimental  Weiss temperatures $\theta_{{\rm p}\alpha}$ in the Curie-Weiss law fitted to magnetic susceptibility data in the PM state of uniaxial single crystals according to~\cite{Johnston2017b}
\be
\theta_{{\rm p}\,ab} - \theta_{{\rm p}\,c} = T_{\rm N}\left(\frac{h_{\rm A1}}{1+h_{\rm A1}}\right),
\ee
where the $ab$~crystal plane corresponds to the $xy$~plane in the theory and the $c$~axis to the $z$~axis, and $T_{\rm N}$ is the measured N\'eel temperature including both exchange and anisotropy contributions.  Then the N\'eel temperature $T_{{\rm N}J}$ due to exchange interactions alone is found from
\be
T_{{\rm N}J} = \frac{T_{\rm N}}{1 + h_{\rm A1}}.
\label{Eq:TNJ}
\ee
The Weiss temperature $\theta_{{\rm p}J}$ in the Curie-Weiss law due to exchange interactions alone is the spherical average
\be
\theta_{{\rm p}J} = \frac{2\theta_{{\rm p}\,ab} + \theta_{{\rm p}\,c}}{3}
\ee
of the measured values $\theta_{{\rm p}\,ab}$ and~$\theta_{{\rm p}\,c}$.

Once $T_{{\rm N}J}$ and $\theta_{{\rm p}J}$ are determined for a particular compound, one can determine the parameters $J_0$, $J_1$, and $J_2$ within the $J_0$-$J_1$-$J_2$ MFT model by solving for them from the three simultaneous equations~\cite{Johnston2015}
\bea
\cos(kd) &=& -\frac{J_1}{4J_2}, \label{Eqs:FindJ0J1J2} \\
\theta_{{\rm p}J} &=& -\frac{S(S+1)}{3}(J_0 + 2 J_1 + 2 J_2),\nonumber \\
T_{{\rm N}J} &=& -\frac{S(S+1)}{3}\big[J_0 + 2 J_1\cos(kd) + 2 J_2\cos(2kd)\big],\nonumber
\eea
where $J_2 > 0$,  the $J_i$ are expressed here in temperature units, and the turn angle $kd$ is assumed to be known from neutron diffraction measurements and/or from fitting the $xy$-plane magnetic susceptibility below $T_{\rm N}$ by MFT~\cite{Johnston2012, Johnston2015, Johnston2015b}.  The solutions for $J_0$, $J_1$, and $J_2$ obtained from Eqs.~(\ref{Eqs:FindJ0J1J2}) are
\bse
\label{Eqs:J012}
\bea
J_0 &=& -\frac{3\csc^4(kd/2)}{8S(S+1)}\Big\{T_{{\rm N}J}\big[1 - 4\cos(kd)\big]\\
&&\hspace{0.9in} +\ \theta_{{\rm p}J} \big[2 + \cos(2kd)\big]\Big\},\nonumber\\
J_1 &=& -\frac{3\csc^4(kd/2)}{4S(S+1)} \big(T_{{\rm N}J} - \theta_{{\rm p}J}\big) \cos(kd),\\
J_2 &=& \frac{3\csc^4(kd/2)}{16S(S+1)} \big(T_{{\rm N}J} - \theta_{{\rm p}J}\big).
\eea
\ese

\subsection{\label{Sec:EuCo2P2} Application to the model molecular-field helical Heisenberg antiferromagnet ${\rm\bf EuCo_2P_2}$}

${\rm EuCo_2P_2}$ is a model MFT helical Heisenberg antiferromagnet with the Eu$^{+2}$ spins situated on a body-centered-tetragonal sublattice with properties given by~\cite{Sangeetha2016}
\bse
\bea
S &=& 7/2,\\
T_{\rm N} &=& 66.6~{\rm K},\\
kd &=& 0.852\pi\,{\rm rad},\label{Eq:kd88}\\
\theta_{{\rm p}\,ab} &=& 23.0~{\rm K},\\
\theta_{{\rm p}\,c} &=& 18.2~{\rm K},\\
H_{\rm c} &\sim& 28~{\rm T}.\label{Eq:HcExtrap}
\eea
\ese
where the value of $kd$ was obtained by neutron diffraction measurements at $T = 15~{\rm K} \ll T_{\rm N}$~\cite{Reehuis92} and the critical field $H_{\rm c}$ is obtained via a long extrapolation of magnetization versus field data at $T=2$~K above the high-field limit $H^{\rm max} = 14$~T of the measurements.  Using $g=2$ and Eqs.~(\ref{Eq:hxTOHx}) and~(\ref{Eq:TNJ})~to~(\ref{Eqs:J012}), one obtains
\bse
\bea
h_{\rm A1} &=& 0.078,\\
T_{{\rm N}J} &=& 61.8~{\rm K},\\
J_0/k_{\rm B} &=& -9.0~{\rm K},\label{Eq:J0val}\\
J_1/k_{\rm B} &=& 1.92~{\rm K},\label{Eq:J1val}\\
J_2/k_{\rm B} &=& 0.54~{\rm K},\label{Eq:J2val}\\
h_{\rm A1}^{**} &=& 11.0 h_{\rm A1} = 0.85,\label{Eq:hA1**}\\
h_x^{**} &=& 32.9 h_x \label{Eq:hx**data}\\
&=& 0.72 H_x[\rm T],\nonumber\\
h_{\rm c}^{**} &=& 32.9 h_{\rm c} \label{Eq:hx**data}\\
&=& 0.72 H_{\rm c}[\rm T],\nonumber
\eea
\ese
where $1~{\rm T} = 10^4$~Oe. The negative value of $J_0$ is consistent with the FM alignment of the moments in each helix layer, and the positive values of $J_1$ and $J_2$ indicate AFM interlayer couplings with $J_2<J_1$ as would be expected.  A positive AFM value of $J_2$ is required to form a helix structure as previously noted.  Using Eqs.~(\ref{Eq:hc**kd>pi/2}) and~(\ref{Eq:hx**data}) and the value of $kd$ in Eq.~(\ref{Eq:kd88}), one obtains predictions for the reduced and actual critical fields as
\bea
h_{\rm c}^{**} &=& 15.6,\\
H_{\rm c} &=& 21.7~{\rm T}.
\eea
The value for $H_{\rm c}$ is seen to be of the same order as the extrapolated experimental value of $\sim28$~T in Eq.~(\ref{Eq:HcExtrap}).

The low-$T$ value $kd = 0.852\pi$~rad for \ecp\ at $T=15$~K  in Eq.~(\ref{Eq:kd88})~\cite{Reehuis92} is closest to the value $kd=3\pi/4$ for the phase diagram in Fig.~\ref{Fig:helix_SF_phase_diagrams}(d), so we compare the experimental data with that phase diagram. The value $h_{\rm A1}^{**} = 0.85$ in Eq.~(\ref{Eq:hA1**}) places \ecp\ near the right edge of this phase diagram where a second-order transition from the $xy$~helix phase to the $xy$~fan phase occurs at a field of approximately one-half of the critical field.  The experimental high-field $ab$-plane magnetization data at temperature $T=2$~K for \ecp\  in Fig.~10 of Ref.~\cite{Sangeetha2016} are in semiquantitative agreement with this prediction, where the experimental value for the weakly first-order $xy$~helix to $xy$~fan {\it crossover} field is $H_{\rm t} \approx 7$~T and the extrapolated critical field $H_{\rm c}$ is estimated as 26~T as discussed above. The differences between the experimental results and the theoretical transiition fields is likely due at least in part to the rather large difference between the observed low-$T$ value  $kd=0.852\pi$~rad and the value $kd=0.75\pi$~rad for which the phase diagram in Fig.~\ref{Fig:helix_SF_phase_diagrams}(d) was constructed.  It also seems likely that the reason the  observed smooth crossover from the $xy$~helix to the $xy$~fan phase is different from the predicted second-order phase transition is because the value of $kd$ in \ecp\ is different from $kd = 3\pi/4$ in Fig.~\ref{Fig:helix_SF_phase_diagrams}(d)~\cite{Johnston2017c}.

\section{\label{Sec:Summary} Summary and Discussion}

The present work is a continuation of the development and use of the unified molecular field theory for systems containing identical crystallographically-equivalent Heisenberg spins~\cite{Johnston2012,Johnston2015,Johnston2015b}.  This MFT has significant advantages over the previous Weiss MFT because it treats collinear and noncollinear AFM structures on the same footing and the variables in the theory are expressed in terms of directly measurable experimental quantities instead of ill-defined molecular-field coupling constants or Heisenberg exchange interactions.

As part of this development, the influences of several types of anisotropies on the magnetic properties of Heisenberg antiferromagnets were calculated~\cite{Johnston2016, Johnston2017, Johnston2017b}, including a classical anisotropy field~\cite{Johnston2017b} that was used to good advantage in the present work.  This allowed the transverse-field dependence of the spin-flop phases of helical antiferromagnets to be easily calculated in the presence of finite XY anistropy.  The present work allowed the possibility of either one or two coexisting spherical elliptical hodographs of the moments in the spin-flop phase that enhanced the flexibility for the system to attain a minimum energy versus applied and anisotropy fields.

Together with the previous work on the $xy$~helix and $xy$~fan phases that occur under $x$-axis fields and their corresponding energies at $T = 0$~\cite{Johnston2017c}, the present results on the spin-flop  and associated fan energies were utilized to construct  $x$-axis field $H_x$ versus anisotropy field~$H_{\rm A1}$  phase diagrams that can be compared directly with low-$T$ experimental magnetization versus transverse field data for helical antiferromagnets.  Care was taken to explain how to do this.  Then a comparison of the theory with the magnetic behavior of the model MFT helical Heisenberg antiferromagnet \ecp\ was carried out, and semiquantitative agreement was found.

Previous theoretical studies have been reported of the helix-to-fan transition at $T=0$ that occurs with increasing $x$-axis magnetic field transverse to the helix $z$ axis when the local moments are confined to the $xy$~plane \cite{Nagamiya1962}.  These authors also calculated the transverse field versus XY anisotropy phase diagram as in our Fig.~\ref{Fig:helix_SF_phase_diagrams} but for small values of the helix turn angle~$kd$ where the moments spin-flop out of the $xy$ plane into a single spherical ellipse phase with the axis of the spherical ellipse parallel to the applied transverse field \cite{Nagamiya1962}.  In the present work the range of $kd$ was extended and the SF phase contained up to two spherical ellipses instead of one.   For $0 < kd < \pi/2$~rad the topology of our phase boundaries and the phases themselves are similar to theirs.  However, we found significant differences between the phase diagram for $kd=3\pi/4$ and the three phase diagrams with $kd<\pi/2$~rad.

Since the theoretical predictions were obtained using MFT, quantum fluctuations are not taken into account and hence the predictions are expected to be most accurate for helical Heisenberg antiferromagnets containing large spins such as Mn$^{+2}$ ions with spin $S=5/2$ and Gd$^{+3}$ and Eu$^{+2}$ ions with $S=7/2$.  Although the calculated phased diagrams are for  $T=0$, in practice this means that experimental data with which the theoretical phase diagrams are compared should include data at temperatures much lower than the AFM ordering (N\'eel) temperature, a restriction that is often easy to accommodate as in the presently-examined case of \ecp.


\acknowledgments

The author is grateful to N.S.~Sangeetha for discussions and collaboration on the model MFT helical Heisenberg antiferromagnet ${\rm EuCo_2P_2}$ that motivated this work.  This research was supported by the U.S. Department of Energy, Office of Basic Energy Sciences, Division of Materials Sciences and Engineering.  Ames Laboratory is operated for the U.S. Department of Energy by Iowa State University under Contract No.~DE-AC02-07CH11358.

\end{document}